\documentclass[a4paper]{jpconf}
\usepackage{graphicx}
\begin{document}
\title{Fully gapped superconductivity in Ni-pnictide superconductors BaNi$_2$As$_2$ and SrNi$_2$P$_2$}

\author{N Kurita$^{1,2}$, F Ronning$^1$, C F Miclea$^1$, Y Tokiwa$^{1,3}$, E D Bauer$^1$,\\ A Subedi$^{4,5}$, D J Singh$^4$, 
H Sakai$^{1,6}$,  J D Thompson$^1$ and \\ R Movshovich$^1$}

\address{$^1$ Los Alamos National Laboratory, Los Alamos, New Mexico 87545, USA}
\address{$^2$ National Institute for Materials Science, Tsukuba, Ibaraki 305-0003, Japan}
\address{$^3$ I. Physikalisches Institut, Georg-August-Universit\"at G\"ottingen, D-37077 G\"ottingen, Germany}
\address{$^4$ Materials Science and Technology Division, Oak Ridge National Laboratory, Oak Ridge, Tennessee 37831-6114, USA} 
\address{$^5$ Department of Physics and Astronomy, University of Tennessee, Knoxville, Tennessee 37996-1200, USA} 
\address{$^6$ Advanced Science Research Center, Japan Atomic Energy Agency, Ibaraki 319-1195, Japan} 

\ead{KURITA.Nobuyuki@nims.go.jp}

\begin{abstract}
We have performed low-temperature specific heat $C$ and thermal conductivity $\kappa$ measurements on the Ni-pnictide superconductors BaNi$_2$As$_2$ ($T_\mathrm{c}$\,=\,0.7\,K) and SrNi$_2$P$_2$ ($T_\mathrm{c}$\,=\,1.4\,K). The temperature dependences $C(T)$ and $\kappa(T)$ of the two compounds are similar to the results of a number of $s$-wave superconductors. Furthermore, the concave field responses of the residual $\kappa$ for BaNi$_2$As$_2$  rules out the presence of nodes on the Fermi surfaces. We postulate that fully gapped superconductivity could be universal for Ni-pnictide superconductors. Specific heat data on Ba$_{0.6}$La$_{0.4}$Ni$_2$As$_2$ shows a mild suppression of $T_\mathrm{c}$ and $H_\mathrm{c2}$ relative to BaNi$_2$As$_2$.
\end{abstract}

\section{Introduction}
Since the discovery of superconductivity in LaFeAs(O,F)\,\cite{Kamihara}, a number of studies have been performed for understanding the origin of high-$T_\mathrm{c}$ in the Fe-pnictide superconductors (SCs). The structure of superconducting gap symmetry, closely related with the paring mechanism, is one of the most important issues to resolve. However, it remains still controversial whether superconductivity in Fe-pnictides is conventional, or  unconventional with or without node(s)\,\cite{Ishida_review}. 
On the other hand, Ni-counterparts with the same crystal structure also superconduct\,\cite{Ronning_review}. Until now, it was reported that some Ni-pnictides had several properties in common with Fe-pnictides, for example, in structural transitions from tetragonal to lower symmetry\,\cite{Ronning_BaNi2As2,Ronning_SrNi2P2}. In contrast, there are crucial differences including the magnitude of $T_\mathrm{c}$ which does not exceed 5\,K in any Ni-pnictide, the absence of magnetism, and a more three dimensional structure of the Fermi surface in Ni-pnictides\,\cite{Ronning_review,Ronning_SrNi2P2}. Therefore, it can be expected that identifying the SC gap symmetry of Ni-pnictides would yield a key information for that of Fe-pnictides.

Here, we present the results of low-temperature thermal transport measurements, as established powerful tools for probing low energy excited quasiparticles, on BaNi$_2$As$_2$ ($T_\mathrm{c}$\,=\,0.7\,K) and SrNi$_2$P$_2$ ($T_\mathrm{c}$\,=\,1.4\,K) which exhibit structural transitions at 130\,K\,\cite{Ronning_BaNi2As2} and at 325\,K\,\cite{Ronning_SrNi2P2}, respectively.  

\section{Experimental details}

Single crystals of Ba$_\mathrm{1-x}$La$_x$Ni$_2$As$_{2}$ ($x=0$ and 0.4 nominal composition) and SrNi$_2$P$_{2}$ were grown in Pb-flux\,\cite{Ronning_BaNi2As2} and Sn-flux\,\cite{Ronning_SrNi2P2}, respectively. Thermal conductivity was measured in a dilution refrigerator on plate-like crystals by a standard one-heater and two-thermometers technique with a heat current $q$\,$\parallel$\,[100]. 
Electrical resistivity was measured for electrical current $J$\,$\parallel$\,[100], using the same crystal with the same electrical contacts as for the thermal conductivity measurements. Specific heat was measured in a dilution refrigerator by a standard heat-pulse method for (Ba,La)Ni$_2$As$_{2}$ and in a PPMS (Quantum Design) by a relaxation method for SrNi$_2$P$_{2}$.
Magnetic field was applied for $H$\,$\parallel$\,[100]. In the thermal conductivity study, we measured two samples from different growths denoted as $^\#$1 and $^\#$2.

%%%%%%%%%%%%%%%%%%%%%%%%%%%%%%%%%%%%%%%%%%%%%%%%%
\begin{figure}
\begin{center}
\includegraphics[width=0.9\linewidth]{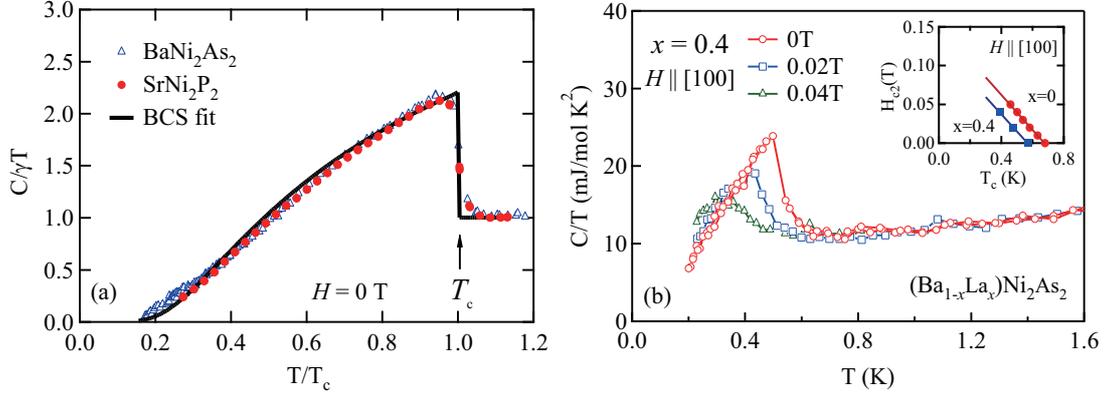}
\end{center}
\caption{(Color online) (a) Scaled $C/T$ vs $T$ of BaNi$_2$As$_{2}$ and SrNi$_2$P$_{2}$ in zero field. Solid curve represents a BCS fit with the parameters found in Ref.\,\cite{KuritaPRL2009}. (b) $C/T$ vs $T$ for Ba$_\mathrm{1-x}$La$_x$Ni$_2$As$_{2}$ ($x=0.4$) in several fields for $H$\,$\parallel $\,$ab$. The inset shows $H_\mathrm{c2}$ vs $T_\mathrm{c}$ in $x=0$\,\cite{KuritaPRL2009} and 0.4 for $H$\,$\parallel $\,$ab$. The solid lines represent a least-squares fit to the data.} \label{fig1}
\end{figure}
%%%%%%%%%%%%%%%%%%%%%%%%%%%%%%%%%%%%%%%%%%%%%%%%%

\section{Results and Discussion}
Figure~\ref{fig1}(a) shows the scaled $C/T$ as a function of $T/T_\mathrm{c}$ of BaNi$_2$As$_{2}$ and SrNi$_2$P$_{2}$ in zero field. 
A nuclear quadrupole contribution arising from As was already subtracted in BaNi$_2$As$_{2}$. 
A theoretical curve based on weak coupling BCS\,\cite{KuritaPRL2009} can be fit to nearly identical data sets for the two compounds. 
This $C/T$ behavior is in strong contrast to the results of nodal SCs such as Sr$_2$RuO$_4$\,\cite{Suzuki_Sr2RuO4}, which has a smaller jump at $T_\mathrm{c}$ as well as much more low energy excitations visible at low temperatures.  
Figure~\ref{fig1}(b) shows $C/T$ vs $T$ of La-doped Ba$_\mathrm{1-x}$La$_x$Ni$_2$As$_{2}$ ($x$\,=\,0.4) in several fields for $H$\,$\parallel$\,$ab$. 
The inset displays the upper critical field $H_\mathrm{c2}$ vs $T_\mathrm{c}$ for $x$\,=0\,\cite{KuritaPRL2009} and $x$\,=\,0.4, determined by the midpoint of the jump in $C/T$. 
The initial slopes $-$0.23\,T/K for $x$\,=0 and $-$0.22\,T/K for $x$\,=\,0.4 give $H_\mathrm{c2}(0)$\,=\,0.11\,T and 0.088\,T, respectively, using $H_\mathrm{c2}$(0)\,=\,$-$0.7\,$T_\mathrm{c}$d$H_\mathrm{c2}$/d$T_\mathrm{c}$\,\cite{WHH}. 
The reduction of $T_\mathrm{c}$ and $H_\mathrm{c2}$(0) by doping in BaNi$_2$As$_{2}$ is opposite to the enhancement of the two values in LaNiAsO by Sr- or F-doping\,\cite{Srdoping,Fdoping}. 
It should be noted that the similar enhacement of $T_\mathrm{c}$ and $H_\mathrm{c2}$(0) with doping was widely confirmed in Fe-pnictides\,\cite{Ishida_review}.

%%%%%%%%%%%%%%%%%%%%%%%%%%%%%%%%%%%%%%%%%%%%%%%%%
\begin{figure}
\begin{center}
\includegraphics[width=0.95\linewidth]{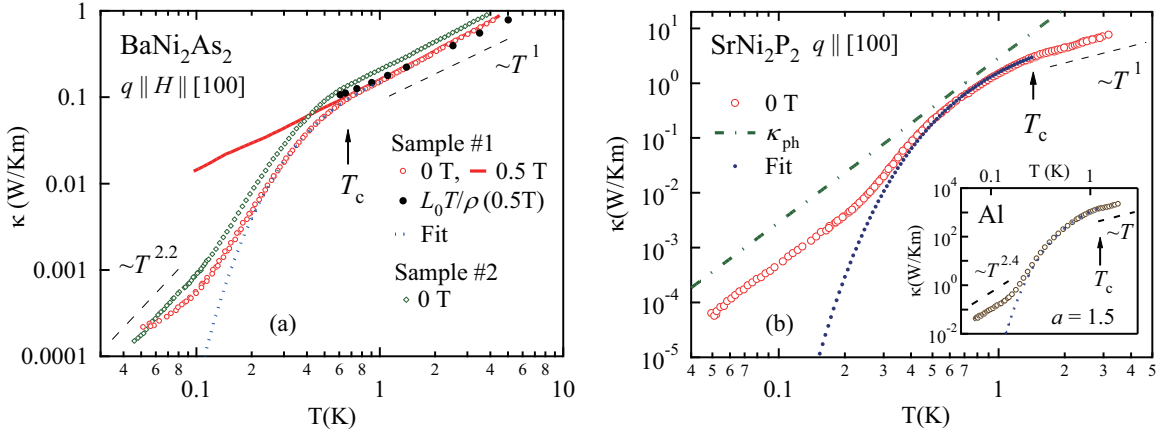}
\end{center}
\caption{(Color online) $\kappa$ vs $T$ in (a) BaNi$_2$As$_{2}$($^{\#}$1\,\cite{KuritaPRL2009}, $^{\#}$2) and (b) SrNi$_2$P$_{2}$ for heat current $q$\,$\parallel$\,[100]. Dotted lines show a fit to $\kappa$($T$) based on BCS theory defined as $\kappa$\,=\,$C$exp(${-a} T_\mathrm{c}$/$T$). Solid symbols in (a) represent the electronic thermal conductivity $\kappa$$_e$\,=\,$L_0$$T$/$\rho$ with $L_0$\,=\,2.45\,$\times$\,10$^{-8}$\,W$\Omega$/K$^2$ for 0.5\,T ($\gg$\,$H_\mathrm{c2}$), derived from resistivity data using the Wiedemann-Franz law. The inset of (b) exhibits $\kappa$ versus $T$ of Al with $T_\mathrm{c}$\,=\,1.2\,K\,\cite{Al_RomanPRL1998}. A dashed-dotted line ($\propto$\,$T^3$) is the upper limit for the phonon conductivity based on scattering from the crystal surfaces.}\label{fig2}
\end{figure}
%%%%%%%%%%%%%%%%%%%%%%%%%%%%%%%%%%%%%%%%%%%%%%%%%

Figure~\ref{fig2} shows the temperature dependence of thermal conductivity $\kappa(T)$ of (a) BaNi$_2$As$_{2}$ ($^{\#}1$\,\cite{KuritaPRL2009} and $^\#2$) and (b) SrNi$_2$P$_{2}$. Heat current $q$ and magnetic field $H$ were applied in the directions $H$\,$\parallel$\,$q$\,$\parallel$\,[100]. For BaNi$_2$As$_{2}$($^{\#}1$), $\kappa$ in zero field above $T_\mathrm{c}$  exhibits approximately $T$-linear behavior and continues to lower temperature in the normal state at 0.5\,T ($\gg$\,$H_\mathrm{c2}$). Using the Wiedemann-Franz law, we estimate the electronic conductivity $\kappa_\mathrm{e}$ in the normal state through $\kappa_e$\,=\,$L_0T/\rho$ with $L_0$\,=\,2.44\,$\times$\,10$^{-8}$\,W$\mathrm{\Omega}$/K$^2$ and the measured resistivity ($\rho$) data at 0.5\,T as indicated by the solid symbols in Fig.~\ref{fig2}(a). The close agreement of the estimated $\kappa_\mathrm{e}$ with the experimental $\kappa$ indicates that the normal state thermal conductivity $\kappa_\mathrm{n}$ can be attributed mostly to electrons rather than phonons. In the superconducting state just below $T_\mathrm{c}$, $\kappa$ exhibits a BCS-like expression exp($-aT_\mathrm{c}$/$T$) as indicated by the dotted curves for BaNi$_2$As$_{2}$ and SrNi$_2$P$_{2}$. It is worth noting that $a$\,=\,1.3 (BaNi$_2$As$_{2}$) and 1.5 (SrNi$_2$P$_{2}$) are comparable to the values 1.3-1.5 obtained in conventional $s$-wave SCs Al, Sn and Zn\,\cite{Berman}. At $T\ll$\,$T_\mathrm{c}$, $\kappa$ does not decrease as rapidly as expected from exponential behavior. Similar deviations from an exponential variation were found in a number of $s$-wave superconductors usually with a form of $\kappa$\,$\propto$\,$T^{\alpha}$ and are commonly attributed with phonon contributions. In fact, $\kappa$\,$\propto$\,$T^{\alpha}$ can be clearly seen in BaNi$_2$As$_{2}$ ($^\#2$) and  SrNi$_2$P$_{2}$ with ${\alpha}$\,=\,2.2 and 3, respectively although an anomalous behavior was found in BaNi$_2$As$_{2}$ ($^\#1$). In particular, ${\alpha}$\,=\,3 in SrNi$_2$P$_2$ is equal to the value expected in a theoretical phonon conductivity $\kappa_\mathrm{ph}$\,$\propto$\,$T^3$, as indicated by a dashed-dotted line\,\cite{Kurita_SrNi2P2} in Fig.~\ref{fig2}(b), based on a crystal-boundary scattering model\,\cite{Berman}. However, there is no consistent explanation why the exponent $\alpha$ ranges between 2 and 3 in many SCs. Note that $\kappa_\mathrm{ph}$\,$\propto$\,$T^{2.2}$ was reported in BaFe$_2$As$_2$ which is a parent compound of Fe-pnictide SCs\,\cite{KuritaPRB2009}. The $\kappa(T)$ behaviors in BaNi$_2$As$_{2}$ and SrNi$_2$P$_{2}$ discussed above are well consistent with the results of a number of $s$-wave superconductors including Al with $T_\mathrm{c}$\,=\,1.2\,K\,\cite{Al_RomanPRL1998} as can be seen in the inset of Fig.~\ref{fig2}(b)

\begin{figure}
\begin{center}
\includegraphics[width=0.95\linewidth]{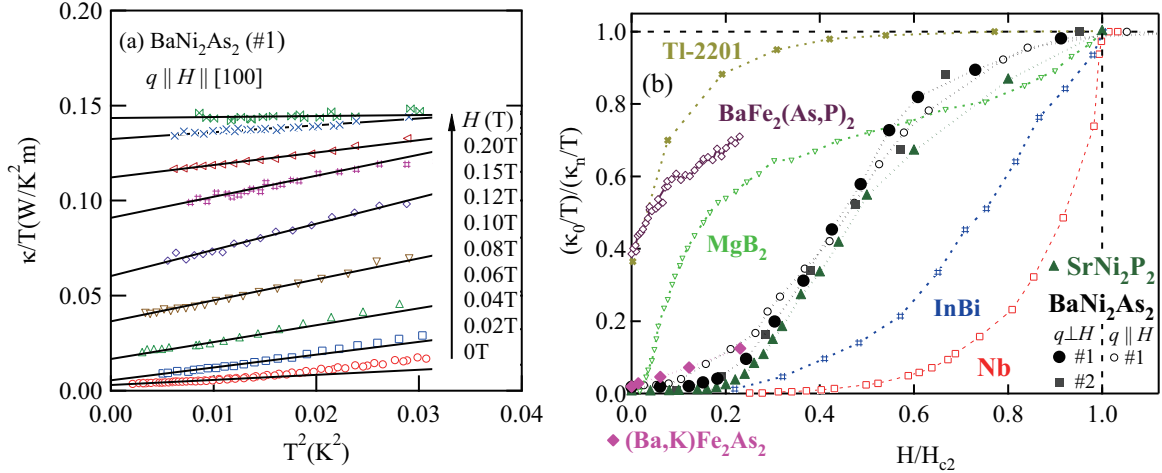}
\end{center}
\caption{(Color online) (a) low-temperature $\kappa/T$ vs $T^2$ of BaNi$_2$As$_2$ in several fields $H$\,$\parallel$\,$q$\,$\parallel$\,[100]. (b) Scaled residual linear term $(\kappa_0$/$T)/(\kappa_\mathrm{n}/T)$ vs $H/H_\mathrm{c2}$ of BaNi$_2$As$_2$ for $^\#1$ ($H$\,$\perp$\,$q$\,\cite{KuritaPRL2009} and $H$\,$\parallel$\,$q$) and $^\#2$ ($H$\,$\parallel$\,$q$), and SrNi$_2$P$_2$ ($H$\,$\perp$\,$q$). Data for Nb (clean, fully gapped $s$-wave)\,\cite{Nb}, InBi (dirty, fully gapped $s$-wave)\,\cite{Wills_InBi}, MgB$_2$ (multi-band gap)\,\cite{Sologubenko_MgB2}, and Tl-2201 ($d$-wave with line nodes)\,\cite{Proust_Tl2201} as well as Fe-pnictide SCs (Ba,K)Fe$_2$As$_2$\,\cite{Luo_BaKFe2As2} and BaFe$_2$(As,P)$_2$\,\cite{Hashimoto_BaFe2AsP2} are shown for comparison.} \label{fig3}
\end{figure}

Figure~\ref{fig3} (a) shows low-temperature $\kappa/T$ vs $T^2$ of BaNi$_2$As$_2$ in several fields $H$\,$\parallel$\,$q$\,$\parallel$\,[100], not presented in Ref\,\,\cite{KuritaPRL2009}. The rise in $\kappa/T$ can only be attributed to $\kappa_\mathrm{e}$, as $\kappa_\mathrm{ph}$ may only go down in magnetic field due to additional scattering from vortices in the mixed state. The straight lines are fits to $\kappa/T$\,=\,$\kappa_0/T$\,+\,$b\,T^2$, where $\kappa_0/T$ is the residual term extrapolated to $T$\,=\,0\,K at each field. Figure~\ref{fig3}(b) displays the normalized $\kappa_0/T$ of BaNi$_2$As$_2$ for $^\#1$ ($H$\,$\perp$\,$q$\,\cite{KuritaPRL2009} and $H$\,$\parallel$\,$q$) and $^\#$2 ($H$\,$\perp$\,$q$), with respect to $H/H_\mathrm{c2}$.  In $s$-wave SCs, quasiparticles localized around the vortex cores at low field ($>$\,$H_\mathrm{c1}$) cannot carry heat until the intervortex distance gets so narrowed with field that quasiparticles can conduct between the vortices. This is the main reason why $\kappa_0/T$ exhibits an exponential-like increase toward $H_\mathrm{c2}$ as seen for Nb data\,\cite{Nb}. In contrast, for nodal SCs such as Tl$_2$Ba$_2$CuO$_{6+\delta}$ (Tl-2201)\,\cite{Proust_Tl2201}, $\kappa_0/T$ exhibits a rapid enhancement in the low field region because of the excitation of nodal quasiparticles by the Volovik effect\,\cite{Volovik}. Therefore, the concave field dependences of $\kappa_0/T$ at low field in BaNi$_2$As$_2$ ($^\#1$ and $^\#2$) and SrNi$_2$P$_{2}$ are clear evidence for fully gapped superconductivity.

The rapid increase of $\kappa$($H$) for BaNi$_2$As$_{2}$ and SrNi$_2$P$_{2}$ more closely resembles the dirty $s$-wave SCs such as InBi\,\cite{Wills_InBi}, than the clean case (i.e., Nb\,\cite{Nb}), which is consistent with the specific heat analysis in BaNi$_2$As$_{2}$\,\cite{KuritaPRL2009}. The shoulder-like anomaly close to $H_\mathrm{c2}$ may reflect a spread in $H_\mathrm{c2}$ as reflected by the width of the heat capacity anomaly in field close to $H_\mathrm{c2}$\,\cite{KuritaPRL2009}. On the other hand, the shoulder might originate from multiband superconductivity although the anomaly is much less dramatic than in MgB$_2$\,\cite{Sologubenko_MgB2}. Further studies should be performed on cleaner crystals to resolve this issue.

As to Fe-pnictide SCs, there is no consistent conclusion for the superconducting pairing symmetry as can be inferred from fully gapped (Ba,K)Fe$_2$As$_2$\,\cite{Luo_BaKFe2As2} and nodal BaFe$_2$(As,P)$_2$\,\cite{Hashimoto_BaFe2AsP2}, both of which have 30\,K-class high-$T_\mathrm{c}$. It is suggested that the strongly $k$-dependent gap function in Fe-pnictides could give rise to nodes on the Fermi surface\,[22$-$24]. 
First principles calculations show that the Fermi surface of the Ni-pnictide compounds is much more complicated than that of the Fe-pnictide compounds because of the two additional valence electrons in Ni relative to Fe\,\cite{Subedi,Singh}. 
Accordingly, it is difficult to find any nodal plane which does not intersect the Fermi surface in the Ni-pnictide compounds.
This leads to the conclusion that the superconducting pairing symmetry of BaNi$_2$As$_{2}$ and SrNi$_2$P$_{2}$ is most likely conventional $s$-wave rather than sign-reversing $\pm s$-wave as suggested in Fe-pnictide SCs\,\cite{Mazin,Kuroki}.

%%%%%%%%%%%%%%%%%%%%%%%%%%%%%%%%%%%%%%%%%%%%%%%%%
%%%%%%%%%%%%%%%%%%%%%%%%%%%%%%%%%%%%%%%%%%%%%%%%%

\section{Conclusion}
To conclude, we have investigated the structure of the superconducting gap symmetry of BaNi$_2$As$_2$ and SrNi$_2$P$_2$ through low-temperature thermal transport measurements. The concave field variations of $\kappa_0/T$ in the low field region have revealed that BaNi$_2$As$_2$ and SrNi$_2$P$_2$ are fully gapped SCs. This conclusion is also supported by the low-temperature dependences of $\kappa$ and $C$.

\section*{Acknowledgment}
We would like to thank I. Vekhter, M. Graf and S.-H. Baek for useful discussions. Work at Los Alamos National Laboratory was performed under the auspices of the US Department of Energy. Work at Oak Ridge was supported by the DOE, Division of Materials Sciences and Engineering.

\end{document}